# STORM: A Model for Sustainably Onboarding Software Testers

Tobias Lorey
University of Innsbruck
Innsbruck, Austria
tobias.lorey@student.uibk.ac.at

Stefan Mohacsi
Atos IT Solutions and Services GmbH
Vienna, Austria
stefan.mohacsi@atos.net

Armin Beer
Beer Test Consulting
Baden, Austria
info@arminbeer.at

Michael Felderer
University of Innsbruck
Innsbruck, Austria
michael.felderer@uibk.ac.at

*Abstract*—Recruiting and onboarding software testing professionals are complex and cost intensive activities. Whether onboarding is successful and sustainable depends on both the employee as well as the organization and is influenced by a number of often highly individual factors. Therefore, we propose the Software Testing Onboarding Model (STORM) for sustainably onboarding software testing professionals based on existing frameworks and models taking into account onboarding processes, sustainability, and test processes. We provide detailed instructions on how to use the model and apply it to real-world onboarding processes in two industrial case studies.

## I. INTRODUCTION

A trend towards integrating *environmental protection*, *social equity*, and *economic development*, also referred to as sustainability [1], has emerged in many industries and academic disciplines in recent years. Examples of this development include sustainable finance [2], corporate sustainability [3], sustainable tourism [4], and sustainable agriculture [5]. Software engineering and software testing are no exceptions to the sustainability trend [6], [7].

Despite being studied in several software engineering domains [8], [9], [10] sustainability has not been used extensively in the context of onboarding software testing professionals. Onboarding *is a procedure whereby employees moving from team outsiders to becoming team members* [11]. It is a cost and labor-intensive activity with the goal to have a newcomer being accepted by the team and begin to work productively.

Experiences from large-scale projects indicate the following problems in onboarding software professionals:

- Onboarding new team members virtually is common practice today, which has even accelerated during the COVID-19 pandemic. However, new team members may miss opportunities to ask questions and establish a closer relationship with their colleagues.
- Employee and company expectations often differ concerning the content and duration of the onboarding phase.
- Onboarding of new employees may be inhibited by not adequately creating a structured onboarding plan.
- Knowledge networking, formal training, and certifications are often not viewed as long-term investments by an organization and lack funding.

As a consequence, the objective of setting up a more sustainable onboarding process has to be considered. Reasons to introduce sustainability in the onboarding process are (1) sustainable and green initiatives gaining interest in industry, (2) increase employee satisfaction and productivity, (3) decrease costs by reducing employee turnover, and (4) increase public reputation as a good place to work.

From the employees' perspective, work-life balance, career opportunities, and job satisfaction play an increasingly important role in their work lives. Such factors are directly related to sustainability. Activities in the onboarding process have to promote these goals to increase employee satisfaction and reduce employee turnover [12].

However, from a company perspective, onboarding activities are limited by budget constraints and the duration of the onboarding phase, given that software testing professionals often need to work productively as soon as possible. Therefore, an organization must find the right balance to incorporate sustainability into existing onboarding processes characterized by organizational and financial aspects and the need to support the test process early on.

Multiple factors impact the sustainability of onboarding new employees into an organization and a team in the context of software testing. Therefore, we see the need to connect the concepts of onboarding and sustainability in the context of software testing professionals. We propose the Software Testing Onboarding Model (STORM) for increasing sustainability in the onboarding process of new software testing professionals, which is structurally based on the floodlight model [13] from the domain of requirements engineering.

This article is structured as follows: In Section 2, we present background and related work on sustainability, onboarding in software engineering and software testing, test processes, and the floodlight model. We describe the proposed Software Testing Onboarding Model and its application in Section 3. Section 4 evaluates the model in two case studies from industry. We discuss the model's application and its implications for industry in Section 5 and provide a conclusion in Section 6.

## II. BACKGROUND AND RELATED WORK

In this section we provide an overview of the concepts of sustainability in software engineering, onboarding new employees, software testing processes, and the floodlight model.

## A. Sustainability in Software Engineering

Penzenstadler & Femmer [14] propose a generic sustainability framework for software engineering that consists of three different levels to improve sustainability in organizations. There are five distinct sustainability dimensions at the top: *individual-, environmental-, economic-, technical-, and social sustainability*. The middle-level consists of *values*, which are morals indicative of each dimension, indicators, and regulations. The bottom-level consists of *activities*, which are specific steps to be taken to achieve the specified goal in the context of sustainability. The framework's practical application consists of two distinct phases: *1) analysis and 2) application & assessment*. The analysis phase consists of instantiating the generic sustainability model, and the application & assessment phase consists of specifying responsibilities and monitoring sustainability using previously defined indicators and metrics. Condori-Fernandez et al. [10] investigate how quality requirements contribute to sustainability dimensions. They found that quality attributes like availability and operability are related to technical sustainability. Kern et al. [15] introduce a causal model for analyzing various sustainability criteria of software. Their focus is on resource and energy efficiency.

## B. Onboarding in Software Engineering and Software Testing

Our work is related to recruiting and onboarding software professionals in organizations and IT projects and incorporating sustainability in the onboarding procedure. We have identified a variety of existing literature on this topic. Begel and Simon [16] focus on the reduction of stress and anxiety during onboarding by fostering social networking for newcomers. They regard strategies like mentoring and pair programming as potential success factors for graduates starting their first jobs. Buchan et al. [11] find that mentoring, online communities, peer support, and team socializing are considered to be the most important onboarding techniques named by professionals in software development companies. They categorize onboarding techniques into the following categories: (1) working with people, (2) working with artifacts, and (3) undertaking an activity. Gregory at al. [17] answer the question of how newcomers should be integrated into an agile project team. They apply Bauer's onboarding framework because it is generic and empirically based.
Caldwell and Peters [18] propose a ten-step model for quality onboarding, observing the organizational impact and employee perception. The goal is to identify ethical implications. Pham et al. [19] found that the lack of testing skills of inexperienced new hires is a problem for software development companies that requires different coping strategies. Florea and Stray [12] investigate educational backgrounds and skill acquisition of software testing professionals. They find that software testers often need to demonstrate curiosity and skills increase with experience. Sharma and Stol [20] explore the relationship between onboarding of new employees and turnover intention. Their research model relates the onboarding activities *orientation, training and support* to onboarding success, which impacts job satisfaction and workplace quality required to reduce turnover intention. The strongest relationship was found to be the impact of support on onboarding success. Brito et al. [21] focus on onboarding activities in large-scale globally distributed projects taking Bauer's model into account. The newcomer's performance is observed in an exploratory case study by evaluating the productivity of the newcomers.

Bauer [22] provides a comprehensive framework for onboarding new employees to an organization. The framework consists of six separate phases. The onboarding process proposed by Bauer is initiated with the *(1) recruiting* of a new employee. After the selection of a candidate the onboarding continues with the *(2) orientation* of the newcomer. The newcomer is introduced to the use of *(3) processes and tools*. *(4) Coaching and support* has to be in place in order to assure efficient onboarding. *(5) Training* aims to improve task performance by practice-based learning, certification and career development. Sensible *(6) feedback* culture and the integration of the newcomer into a team facilitates the onboarding process.

## C. Docker's Floodlight Model

Docker [13] proposed a model to highlight the interactions between requirements and acceptance criteria in the context of requirements engineering. The floodlight model explains how requirements define one or more possible solutions to a problem. These potential solutions are then filtered and limited by predefined acceptance criteria and other constraints until acceptable solutions are found. The name floodlight model is used to visualize how various acceptance criteria cast floodlights on possible solutions. Its terminology is inspired by floodlights used on a theater stage. Having multiple floodlights overlap on a solution implies that the criteria are fulfilled.

## III. STORM: SOFTWARE TESTING ONBOARDING MODEL

This section describes the Software Testing Onboarding Model (STORM) which takes into account emerging sustainability factors. We divide this section into a static and dynamic view. The former describes the model's concepts and their relationships among each other while the latter provides guidance on the application steps.

### A. Constructs

The model's structure is based on the floodlight model by Docker [13]. We adapt and amend Docker's floodlight model to the context of onboarding software testers to identify suitable onboarding solutions which comply with the onboarding process, sustainability dimensions, the test process, and further relevant criteria of companies and projects regarding organizational and financial aspects.

The following constructs are part of the framework as shown in Figure 1.

*Onboarding Process:* The onboarding process defines the problem space of our model. We refer to the phases of a standard onboarding process [22] (recruiting, orientation, processes and tools, coaching and support, training, and feedback) which represent sources of problems that must be solved for sustainably onboarding software testing professionals.

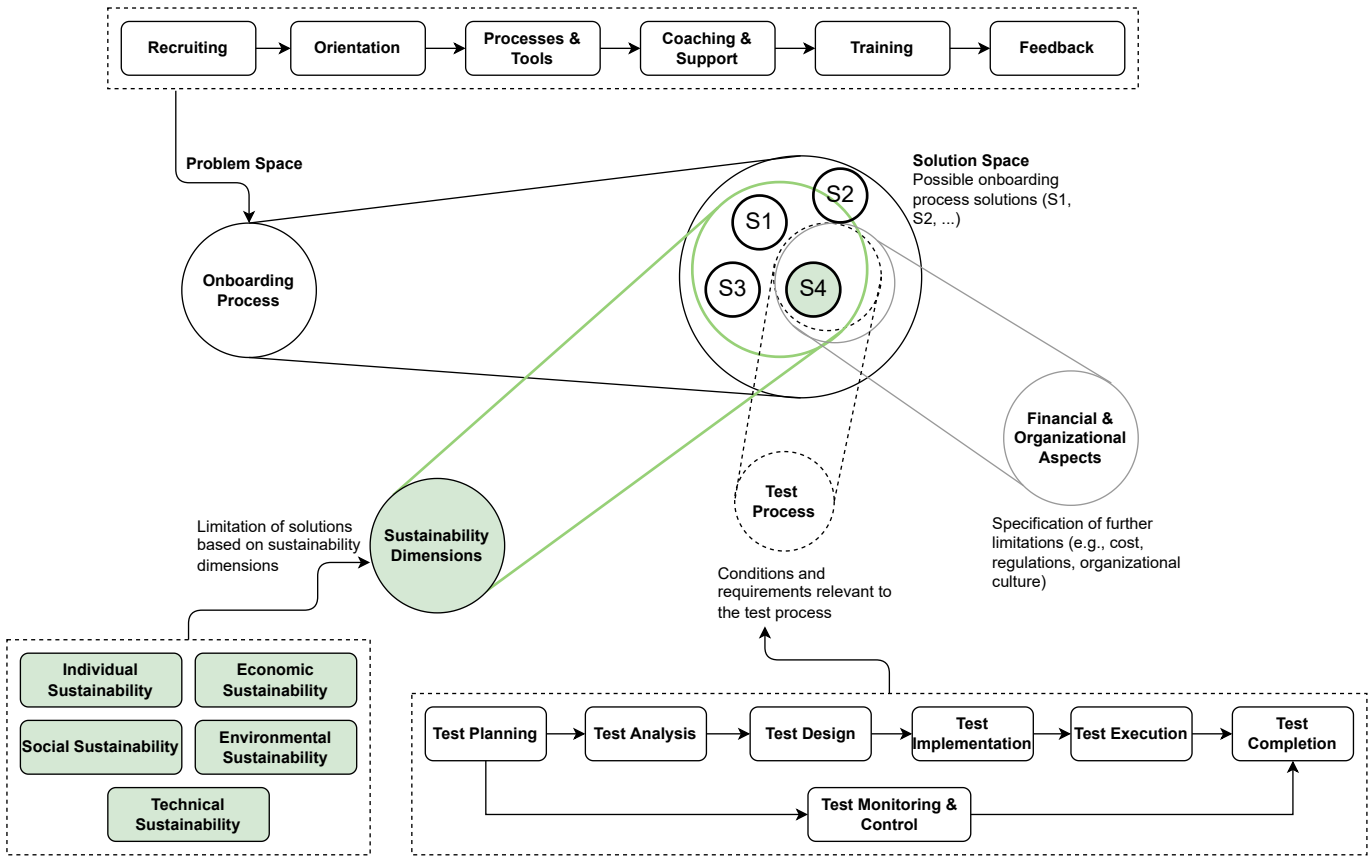

Fig. 1. STORM: Software Testing Onboarding Model

*Solution Space:* A set of possible onboarding solutions (S1, S2, S3, ...) is defined to address identified challenges in the onboarding process of a software testing professional. Potential onboarding solutions are subsequently analyzed regarding their sustainability fit, test process fit, organizational fit, and financial aspects. To visualize the interaction between these aspects and solutions, each of the floodlights cast a light on acceptable onboarding solutions. Overlapping floodlights indicate acceptable solutions regarding multiple acceptance criteria. The solution space is similar to the solutions as introduced by Docker [13] in the original floodlight model in the context of requirements engineering but are of a more proactive nature that reflects the onboarding process consisting of phases and activities as well.

*Sustainability Dimensions:* Potential onboarding process solutions and activities are screened for their sustainability fit. Sustainability dimensions of interest in the software engineering domain include individual, economic, social, environmental, and technical sustainability [14]. Depending on the company's goals, some sustainability dimensions may be considered more important than others or may even be obsolete. These dimensions can be relevant for the company, its employees and stakeholders, the environment, society as a whole or a community, or any combination of the above.

*Test Process:* Onboarding solutions for software testers must support the current or desired test process. They must, therefore, fit the organization's test process in regard to aspects such as training and certifications, agile practices, project structure, and testing tools. Solutions can be analyzed and filtered based on the existing test process or a target test process. A standard ISTQB [23] test process includes the phases *test planning, test monitoring and control, test analysis, test design, test implementation, test execution, and test completion* and can be used to identify whether onboarding solutions, e.g., certifications and trainings, fit the test process.

*Financial & Organizational Aspects:* Solutions are further examined in regard to their organizational and financial aspects, e.g., implementation cost, legal and regulatory requirements, organizational processes and culture.

### B. Model Application

This section describes the application of the onboarding model as shown in Figure 2. The model is designed to be incorporated into existing onboarding processes and serves as a tool for both designing and analyzing onboarding processes focused on the software testing domain.

We divide the onboarding process into three distinct phases: preparation and recruiting phase, application of the Software

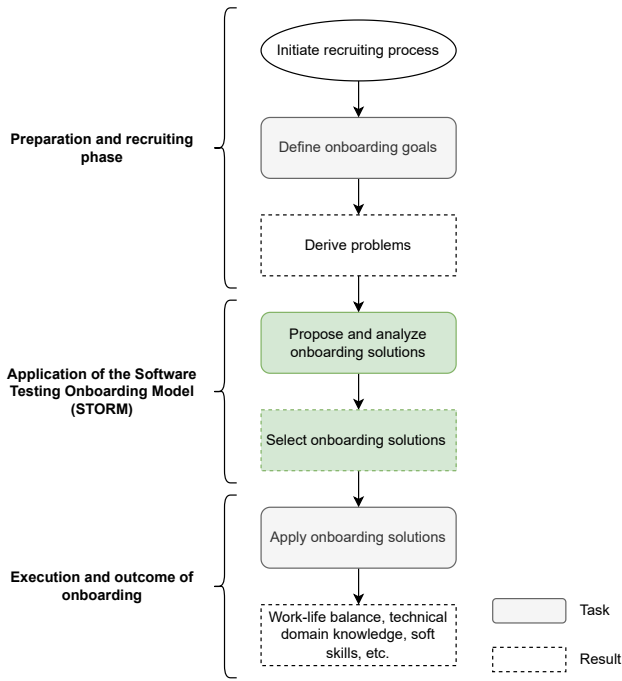

Fig. 2. Onboarding Model: Application

Testing Onboarding Model and execution and outcome of the onboarding process. The preparation phase consists of initiating the recruiting of a suitable software testing professional, as well as defining specific onboarding goals from which problems are derived. Solutions need to be proposed to solve these onboarding challenges. The onboarding model is applied at this stage to facilitate analysis of solutions and filter acceptable ones. Each potential solution is analyzed step-by-step in regard to (1) sustainability, (2) testing processes, and (3) organizational constraints such as financial aspects. Each step of the analysis classifies a solution whether it is *positive, neutral,* or *negative* in the respective category. More granular scales, e.g., a Likert scale, may be used if required. Depending on the outcome of the analyses, a set of solutions and activities will be applied in the onboarding process. Selected acceptable solutions will be implemented during the onboarding phase of a new software testing employee. The outcome should be closely monitored in regard to the actual impact of the executed solutions.

## IV. Industrial Cases

We will now apply the onboarding model described in Section 3 to two industrial case studies in an Austrian social insurance institution and a global IT services company (Atos). The two presented companies are developing software applications. Both organizations require skilled software testing professionals to work in IT projects and face various challenges during the recruiting and onboarding phase of newcomers.

### A. Industrial Case 1: Social Insurance Institution

The Austrian social insurance institution offers a wide range of medical services to its insurants. It consists mainly of a department for developing internal software and business units for managing the medical care of about one million insurants. Software development is performed co-located by internal and external personnel. This case study focuses on hiring and onboarding a test automation engineer. The onboarding model was applied by test management experts of the social insurance institution. Problem statements and possible solutions are summarized in Figure 3.

**Onboarding Goals:** Onboarding a new testing professional in the social insurance institution closely interacts with the company's environmental, economic and organizational aspects. The career of a new software testing employee depends on the individual skill level: A junior tester must have the ability to perform the assigned work packages according to the guidelines and defined processes. A junior tester should hold a Foundation Level certificate of the ISTQB. By designing test cases based on the requirements specification, the tester must communicate with analysts, project managers, and software developers in a project. The tester should also be able to pass a software testing curriculum. Requirements for a senior tester are more focused and include technical competence, soft skills, and in-depth knowledge of requirements. A senior testing professional can identify problems and suggest improvements in test case design, development, and release management.

**Onboarding problem 1: Which person shall I hire?** In all projects, iterative and incremental development and a standard test process were used. Automated functional and performance tests were implemented in the majority of projects. However, the requirements coverage was too low, and the effort and duration of maintaining the test cases and test environment made the process inefficient. Therefore, the social insurance institution decided to hire new testers to satisfy the demand for new or ongoing projects. The typical path of integrating a novice tester into a project begins by building testing knowledge and continues with the domain knowledge of the business unit. In addition, soft skills are needed, for instance, to report test results to the stakeholders.

**Possible solutions and results:**
- Solution 1 – A manual tester with the skills of a junior is cheaper than a test automation engineer. Experience from different projects in the social insurance institution showed that their skills are not sufficient to develop an economically efficient test environment.
- Solution 2 – An employee of a business unit with sound domain knowledge can be trained in workshops to learn systematic test case design methods. The newcomer will be classified as a *junior tester* in the organization's career path. Nevertheless, additional time during onboarding needs to be provided. This is a viable and economically sustainable solution.
- Solution 3 – Hiring an external test automation specialist is expensive but recommended to put a framework into operation and enable organizational learning in the team. However, it must be guaranteed that the specialist will be available for the entire project duration of three years to ensure sustainability.

| Problem | Possible Solutions | F1 – Pathways to Sustainability | F2 – Test Process | F3 – Financial and Organizational Aspects | Result |
|---|---|---|---|---|---|
| P1 – Which person shall I hire? | P1S1 – Hire manual tester | Manual testing requires more effort in the long term and is therefore not economically sustainable | The current test process focuses mainly on manual testing | Manual tester is cheaper than an automation specialist | N |
| | P1S2 – Nominate employee of internal business unit | Employee with good domain knowledge requires effort to train methods of test case design | Test case design benefits from bridging effective requirements management and testing | Additional time for courses and coaching in test case design | (Y) |
| | P1S3 – Hire external test automation engineer | Onboarding a test automation engineer creates value and therefore economically sustainable | The current test process needs to be adapted to enable test automation | Test automation specialist is more expensive, but affordable | (Y) |
| P2 – How could I alleviate the task of a test automation engineer? | P2S1 – Attain ISTQB Certification | Certification in CTFL, ALTM and test automation is the basis | The test process benefits from systematic testing | Budget has to be planned | (Y) |
| | P2S2 – Review of testability of requirements specification | Better insight of requirements specification | The test automation benefits from improved testability | Additional cost of improvement of requirements specification | (Y) |
| | P2S3 – Reduce the number of automated test cases | Shift share of automated to manual test cases | Risk of bad quality of releases | Added budget for correction and retest | N |
| | P2S4 - Internal TOSCA training | Experience facilitates ecological thinking | The test process benefits from experience of successful automation with TOSCA | Resources and time needed for effective collaboration | (Y) |
| | P2S5 - TOSCA training by external organization | Consulting and TOSCA certificate facilitate social sustainability | Test process benefits from experienced consultants of tool manufacturers | Budget for support inhibits sufficient consulting. | N |

| N | Negative - should not be implemented |
| (Y) | Neutral - considered for implementation |
| Y | Positive - should be implemented |

Fig. 3. STORM application in industrial case 1: Social insurance institution

**Onboarding problem 2 – How could I alleviate the task of a test automation engineer?** According to the ISTQB glossary, the role of a test automation engineer is *a person responsible for the design, implementation, and maintenance of a test automation architecture as well as the technical evolution of the resulting test automation solutions* [23]. At the social insurance institution, the critical goals of a new test automation engineer were defining a test automation strategy and sharing assets and methods to ensure their consistent implementation across the organization.

Transitions of co-located teams into virtual teams are nowadays common in industry [24]. In the social insurance institution, relationships among team members and spontaneous opportunities to learn skills on the job were important. However, the spatial separation made it more difficult to innovate testing procedures and follow improvements.

**Possible solutions and results:**

- Solution 1 – Professional certification is viewed as the basis for domain and technical knowledge and the career path of a new employee. A new test automation engineer should attain at least the ISTQB Foundation Level certificate, one or more Advanced Level certificates (most notably the Technical Test Analyst), and possibly even the Expert Level of test automation. The application of this solution depends on the budget for workshops and certification exams.
- Solution 2 – Test automation benefits from the high testability of the requirements. A newcomer should review

the testability of the requirements to get an insight into his tasks [25]. Improving the requirements specifications pays off concerning increased efficiency in release planning and better maintainability of test cases despite its cost.

- Solution 3 – Another option is to prioritize and reduce the number of automated test cases. As part of the iterative-incremental development process, the question of which new features should be incorporated in a release is vital for product success. Quality-driven resource planning must be applied for all available resources [26]. To prioritize and reduce the number of automated test cases is no option as it would substantially lower test efficiency and effectiveness.
- Solution 4 – Internal testing software training by employees of a successful project and domain experts is a cost-efficient way to train a new employee and training sessions can be conducted both on premise as well as virtually.
- Solution 5 – Training and consulting by the testing tool manufacturer is another way of know-how transfer for a test automation engineer. However, experience from previous workshops in the social insurance institution indicates that this solution is not sustainable, especially concerning its cost.

*B. Industrial Case 2: Large IT Service Provider (Atos)*

Atos is a global IT services company with over 100,000 employees across 72 countries. Atos is offering a wide range of services including test consulting, test management, and test automation. The following use case was performed by test experts of the Austrian Digital Assurance department who applied the Floodlight Model to a couple of non-trivial problems. Problem statements and possible solutions are summarized in Figure 4.

**Onboarding Goals:** When onboarding new testers, Atos focuses on technical, social, and economic sustainability. Firstly, the hiring strategy is not limited to tools and technologies that are currently in use but also considers upcoming trends. Another important goal is integrating newcomers quickly into the team, which gives them the opportunity to establish social ties and raises productivity.

**Onboarding problem 1 – Which testing skills are the most relevant for a job candidate?** Depending on the customer and the circumstances of the project, different testing approaches are being used. If the technical prerequisites for test automation are not fulfilled or the customer process is too chaotic (e.g., frequent changes of the user interface), manual test execution is still applied.

For all other projects, test automation is strongly encouraged. However, the question whether to use a traditional, well-established automation technique or to introduce a new, innovative approach needs to be resolved.

The decision is also crucial for the hiring strategy. The skill set for new testers has to correspond to the testing approach in the upcoming projects.

**Possible solutions and results:**

- Solution 1 – Manual testers are usually much cheaper than test automation engineers. Learning the theoretical background of SW testing theory and obtaining the ISTQB Foundation Level certification is only a matter of a few weeks and requires little technical skills or previous experience. On the other hand, hiring manual testers is less economically sustainable because they cannot be used for other projects in which test automation is applied.
- Solution 2 – The existing test automation approach focuses on traditional GUI-based test automation using commercial tools that simulate user actions and verify the outcome on the screen. This method is well-established and integrated into the existing test process. A test automation engineer who is familiar with this approach is more expensive than a manual tester but also more flexible due to their technical background. However, there remains the question if the traditional automation approach is fit for the future.
- Solution 3 – Innovative test automation approaches such as AI-based testing and comprehensive RPA with tools such as UiPath have been emerging in recent years. Hiring test automation engineers who are familiar with these new techniques is tempting, but first the additional cost and effort for adapting the test process, buying new tool licenses etc. need to be considered. In the long term, investing in new technologies is more sustainable, but in our opinion the transition should be done incrementally rather than with a radical disruption. Thus, we decided to try out some innovative approaches in the scope of pilot projects while sticking to the well-established method in most other projects. The hiring was done accordingly – a number of young test automation engineers who already had some experience with the innovative methods was hired as well as some engineers who were familiar with the traditional approach.

**Onboarding problem 2 – How can new employees be integrated in a team despite COVID-19 restrictions?** Integrating a new employee into an existing project team can be difficult at the best of times but has been particularly challenging during the COVID-19 crisis. With personal meetings prohibited or severely restricted, most employees had to work exclusively from their home office. A get-together with other team members was only possible in the form of online meetings, which despite the use of video cameras cannot fully replace a personal meeting.

Nonetheless, the switch to exclusive home office went very smoothly at Atos. Most employees were already used to occasionally working in home office and already had the technical equipment in place. Also, at the beginning of the pandemic there was a notable wave of solidarity and team spirit amongst the employees who had realized that they could only master the crisis together. The team members already knew each other well and social contacts between them were well-established.

| Problem | Possible Solutions | F1 – Pathways to Sustainability | F2 – Test Process | F3 – Financial and Organizational Aspects | Result |
|---|---|---|---|---|---|
| P1 – Which testing skills are the most relevant for a job candidate? | P1S1 – Manual testing | Manual testing requires more effort in the long term and is therefore not economically sustainable | Manual testing is still being used in some projects | A manual tester is cheaper than an automation specialist | N |
| | P1S2 – Experience with existing test automation tools | More economically sustainable than manual testing but less than an innovative approach | The current test process is adequate for the existing tool framework | An automation specialist is more expensive. No extra tool licenses required. | (Y) |
| | P1S3 – Experience with innovative test automation tools | Investing in innovative trends is more economically sustainable in the long term. | The current test process needs to be adapted to the new tool framework | An automation specialist is more expensive. New tool licenses required. | (Y) |
| P2 – How can new employees be integrated into a team despite COVID-19 restrictions? | P2S1 – Direct assignment to customer project | Having to work with unfamiliar people is a danger to social sustainability | Collaboration with other team members might be difficult | No additional costs | N |
| | P2S2 – Online team building event | Getting to know the other team members is important for social sustainability | Collaboration with other team members is strengthened | A team event requires time and money | (Y) |
| | P2S3 – Daily online team meetings | Daily contact with the other team members improves social sustainability | In an Agile project, daily standup meetings are part of the culture. | No additional costs | Y |
| | P2S4 – Mentoring | Being introduced by a mentor into the team can improve social sustainability | Faster integration into the team and the project | Some extra budget is required for mentor and mentee | (Y) |

| N | Negative - should not be implemented |
| (Y) | Neutral - considered for implementation |
| Y | Positive - should be implemented |

Fig. 4. STORM application in industrial case 2: IT services company (Atos)

However, this was not the case for new employees who had to be integrated into the existing teams. The need for new hirings was rather intensified than abated by the pandemic due to an increased number of demands by government agencies. But how can new employees be welcomed into a team and social sustainability be achieved if personal meetings are impossible?

**Possible solutions and results:**

- Solution 1 – The simplest option is directly assigning new employees to a customer project and hoping that they will get to know the other colleagues in the course of time. This will generate no additional costs but might make the collaboration with other team members more difficult. If we put the focus on social sustainability, this option can be clearly dismissed due to the danger of new employees feeling uncomfortable because they have to work with people they've never met.
- Solution 2 – An online team building event in which the new employees are introduced to the other team members and get the opportunity to establish social ties can improve social sustainability and future collaboration. On the other hand, extra time and budget is required. Thus, we decided to organize occasional events for welcoming several new employees at a time, but not a separate event for every new staff member.
- Solution 3 – In agile projects, daily stand-up meetings have been a key part of the culture even before the pandemic. During the COVID-19 crisis, these meetings proved invaluable to maintain social contacts between the team members and avoid loneliness. Apart from discussing professional topics, the regular online meetings at 9am gave old and new employees alike the opportunity to talk about their feelings of uncertainty, grief, and fear that were caused by the pandemic. In our experience, connecting with people on an emotional basis has been the most effective stimulus for social sustainability.
- Solution 4 – Being introduced by a mentor to the other team members and into the project is a good way to facilitate collaboration and improve social sustainability. Some extra time and budget are required for mentor and mentee, but we found that this investment usually pays off very quickly.

## V. DISCUSSION

We successfully applied the Software Testing Onboarding Model (STORM) to two industrial cases. The onboarding model enabled company-specific suggestions for relevant onboarding problems for which we received positive feedback from the experts who evaluated the Software Testing Onboarding Model.

If a test manager in the social insurance institution considers hiring a new employee, the software testing onboarding model facilitates the effective integration into a team according to the expert. It promotes career paths for new employees as well as pathways towards sustainability and overall test process improvement. The same is valid for the IT services company Atos. Our model enabled the expert to analyze a total of seven possible solutions to two relevant onboarding problems. For instance, daily online team meetings fit the agile culture of the company, improve social sustainability through regular discussions, and can be introduced at no additional cost.

While the onboarding problems were specific to the two organizations, we believe that these onboarding problems can be generalized to other companies in the IT sector as many face similar situations.

## VI. CONCLUSION

In this article, we proposed a model for sustainable onboarding in the context of software testing. We described the structure of the model and the relationship among its constructs. The potential benefits of the implementation of the presented onboarding model STORM in an organization are:

- contribute an onboarding solution that integrates into the onboarding and software lifecycle of different types of companies and projects,
- gain a better understanding of the organizational environment, infrastructure, and test process related to sustainable onboarding,
- design and develop (new or modified) onboarding solutions that meet sustainability needs for both employees and companies,
- assess and mitigate risks during the onboarding of new employees,
- verify and improve onboarding solutions.

We demonstrated its applicability in two industrial case studies. In the future we plan to refine the proposed onboarding model and distinguish between onboarding in co-located, distributed, hybrid, and on-premise organizations.